\documentclass[a4paper,11pt]{article}
\usepackage{epsf}

\title{Short--time rotational diffusion in monodisperse
       charge--stabilized colloidal suspensions\footnote{
          Dedicated to Prof. Rudolf Klein on the occasion
          of his 60th birthday
       }
   }
\author{M. Watzlawek and G. N\"agele \\ \hfill \\
           {\small Fakult\"at f\"ur Physik, Universit\"at Konstanz,} \\
           {\small Postfach 5560,
           D--78434 Konstanz, Germany}}
\date{{\small (Accepted for publication in Physica A)}}

\begin{document}

\maketitle

\begin{abstract}
        We investigate the combined effects of electrostatic interactions
        and hydrodynamic interactions (HI) on the short--time rotational 
        self--diffusion
        coefficient $D_s^r$ in charge--stabilized suspensions.
        We calculate $D_s^r$
        as a function of volume fraction $\phi$ for various effective particle
        charges and various amounts of added electrolyte.
        The influence of HI is
        taken into account by a series expansion of the two-body mobility 
        tensors.
	At sufficiently small $\phi$ this is an excellent 	
	approximation due to the strong electrostatic repulsion.
	For larger $\phi$, we also consider the leading 
        hydrodynamic three--body 
	contribution. 

	Our calculations show that 
	the influence of the HI on $D_s^r$ is less pronounced 
        for charged particles 
	than for uncharged ones. Salt--free suspensions are 
        particularly weakly influenced by HI.
	For these strongly correlated systems we obtain the 
        interesting result 
	$D_s^r=D_0^r(1-a_r\phi^2)$ 
	for small $\phi$. 
	Here $D_0^r$ denotes the Stokesian rotational diffusion coefficient, 
        and
	$a_r$ is a positive 
	parameter which is found to be nearly independent of the particle 
        charge.
	The quadratic $\phi$--dependence can be well explained in terms 
        of an effective 
	hard sphere model.

	Experimental verification of our theoretical results for $D^r_s$
	is possible using depolarized dynamic 
	light scattering 
	from dispersions of optically anisotropic spherical particles. 
\end{abstract}

\section{Introduction}

Until several years ago, the major interest in the dynamics of colloidal
suspensions was focussed on the investigation of translational diffusion.
The properties of translational collective and self--diffusion have been
studied in great detail both experimentally \cite{Pusey:91} and
theoretically 
\cite{Hess:Klein:83,Jones:Pusey:91,Naegele:Habil:published}.
Information on the translational diffusion is contained in the dynamic 
structure
factor $S(q,t)$, which can be probed for a certain range of wavenumbers 
{\boldmath$q$} and correlation
times $t$ using dynamic light scattering from 
essentially monodisperse particles
\cite{Pusey:91}.

More recently, considerable effort was made to investigate also the rotational
diffusion in suspensions of spherically shaped colloidal particles, which 
possess 
an intrinsic optical anisotropy due to a partially crystalline structure
\cite{Degiorgio:94,Degiorgio:95}. To investigate the rotational particle
diffusion, one has to resort to depolarized dynamic light scattering
(DDLS), which is sensitive on the temporal self--correlations both
in the particle positions and orientations. With this technique, 
the information on the
short--time rotational diffusion is obtained from the 
autocorrelation function $G_{VH}(q,t)$ of the horizontally polarized scattered
electric field \cite{Berne:Pecora}. From DDLS measurements of the first
cumulant of $G_{VH}(q,t)$, one can determine the short--time
rotational self--diffusion coefficient $D^r_s$. The diffusion coefficient
$D^r_s$ depends on the direct potential interactions 
between the colloidal particles and on the indirect 
hydrodynamic
interactions (HI) mediated by the suspending solvent.
The latter interactions account for the fact that
the velocity field, generated in the 
surrounding fluid by the motion of one particle, affects that of the
other particles.

By performing DDLS on index--matched suspensions of charged spherical 
par\-ticles made of
a fluorinated polymer, Degiorgio et. al. were able to measure in great
detail the concentration dependence of the rotational diffusion coefficient
\cite{Degiorgio:94,Degiorgio:95}. They were mainly interested in 
suspensions with
large salinity, where the particles essentially behave as hard spheres. Their
experimental findings for $D^r_s$ have been compared with recent theoretical
calculations by Jones \cite{Degiorgio:95,Jones:1:88}, designed for 
hard sphere suspensions.
These calculations include also the approximative evaluation of the coefficient
of the quadratic term in the virial expansion of $D^r_s$ in powers of 
the volume fraction
$\phi$. The agreement between theoretical and experimental results for 
$D^r_s$ is
found to be satisfactory. Moreover, the small differences observed in the
magnitude of the first and second virial coefficients are suggested to 
be partially
due to residual electrostatic interactions, which are left over if the ionic
strength of the added electrolyte is not large enough to completely screen the
Coulomb repulsion \cite{Degiorgio:95}.

In fact, the microstructure of a suspension of charged particles depends 
crucially
on the amount of added electrolyte. For charge--stabilized suspensions, the
general form of, e.g., the pair distribution function $g(r)$ is qualitatively 
different from that
of hard spheres at the same volume fraction \cite{Naegele:Habil:published}.
Due to the large spatial extend of the counterion cloud, the pair 
potential can have a range
much larger than the physical particle diameter $\sigma$. Therefore, 
appreciable pair 
correlations become already important at much lower values of the 
volume fraction $\phi$
as in the case of hard spheres. For this reason, the concentration dependence
of $D^r_s$ can be quite different for uncharged and charged particles, 
particularly
if the amount of added electrolyte becomes low.

In this work, we focus on the combined effects of electrostatic and 
hydrodynamic
interactions on the short--time rotational self--diffusion in 
charge--stabilized suspensions
of spherical particles. Starting from the generalized Smoluchowski 
equation for the positional
and orientational degrees of freedom, we calculate $D^r_s$ as a 
function of volume fraction
$\phi$ for various effective particle charges and various amounts 
of added electrolyte.
The hydrodynamic interactions are taken into account by a series 
expansion of the two--body
mobility tensors in terms of the reciprocal interparticle distance, 
$r^{-1}$, by including
contributions up to order $r^{-20}$. 
At sufficiently small $\phi$, this in an excellent approximation 
due to the strong
electrostatic repulsion, which renders configurations of nearly 
touching particles
as extremely unlikely. For larger $\phi$, we consider also the 
leading hydrodynamic 
three--body contribution \cite{Degiorgio:95}, using Kirkwood's 
superposition 
approximation for the triplet correlation function, which is
needed as the static input.

We will show that the influence of the HI on $D^r_s$ is less strong for
charge--stabilized particles than for uncharged ones. The influence of HI is 
particularly weak for suspensions in which all exess ions have been removed.
For these strongly correlated systems, we find the remarkable result
$D_s^r=D_0^r(1-a_r\phi^2)$ for small $\phi$, where $D^r_0$ is the 
rotational diffusion coefficient at infinite dilution. The parameter $a_r$ is 
found to be nearly independent of the particle charge for values 
large enough so that
the hard core is completely masked by the electrostatic repulsion.
We can explain the quadratic $\phi$--dependence in terms of a 
simplified calculation
based on an effective hard sphere model.

The paper is organized as follows. In Sec. \ref{ddlssection}, we
briefly summarize salient relations of the theory of DDLS from optically
anisotropic particles, which are relevant for obtaining the first cumulant 
of the 
depolarized field autocorrelation function. For later comparison, we recall in
Sec. \ref{diffusionsection} the theory of short--time 
rotational diffusion and its
main results with regard to suspensions of hard spheres.
Sec. \ref{modelsection} gives the details of our calculations of 
$D^r_s$ for charge--stabilized suspensions. 
It contains also a discussion of the range of
validity of the various approximations used in this work.
We present our results in Sec. \ref{resultssection} and discuss their meaning 
in terms of simple physical arguments based on an effective hard sphere model.
Sec. \ref{conclusionsection} contains our final conclusion.

\section{Depolarized light scattering from anisotropic particles}
\label{ddlssection}

In this section, we summarize some pertinent relations of the theory 
of depolarized
dynamic light scattering (DDLS) from optically anisotropic particles, which are
required for our further discussion. A thorough discussion of general DDLS
properties has been given in \cite{Degiorgio:94,Degiorgio:95,Berne:Pecora}, 
which
allows us to be rather brief in our explanations. 

Consider a suspension of identical spherical particles with cylindrically 
symmetrical optical 
anisotopy and in a situation in which the incident electric 
field of the laser beam is 
linearly polarized perpendicular to the scattering plane. 
The suspended particles are
assumed to be nearly index--matched to the solvent so that 
the Rayleigh--Gans--Debye
approximation is valid even for large volume fractions. The 
presence of optical anisotropy
in the scatterers gives then rise to a non--vanishing horizontally polarized 
part of the scattered electric field with magnitude
\begin{equation}
   E_{VH}(\mbox{\boldmath$q$},t)=
      f(q)\beta\left(\frac{2\pi}{15}\right)^\frac{1}{2}\sum_{l=1}^N
      \left(
         Y_{2,1}(\mbox{\boldmath$u$}_l(t))+
         Y_{2,-1}(\mbox{\boldmath$u$}_l(t)) 
      \right)e^{i\mbox{\boldmath\scriptsize$q$}\cdot
                \mbox{\boldmath\scriptsize$R$}_l(t)},
\end{equation}
which is not affected by multiply scattered light. Here $N$ is the 
number of particles in the
scattering volume,
{\boldmath$q$} is the scattering vector with modulus $q$, and 
$\mbox{\boldmath$R$}_l(t)$ and
$\mbox{\boldmath$u$}_l(t)$ are the position vector and, respectivly,
the unit vector in the
direction of the optical axis of the $l$th particle.
The $Y_{2,\pm 1}$ are the second--order spherical harmonics of index $\pm 1$, 
$f(q)$ is the 
scattering amplitude of a particle, and 
$\beta=\alpha_\parallel -\alpha_\perp$, is the
internal particle anisotropy, i.e. the difference in the 
polarizabilities parallel and
perpendicular to the optical axis.

In DDLS experiments, one measures the
modulus of the temporal correlation function \cite{Degiorgio:94,Degiorgio:95}
\begin{equation}
   G_{VH}(q,t)=\langle E_{VH}(\mbox{\boldmath$q$},0)
                       E^*_{VH}(\mbox{\boldmath$q$},t)\rangle
\end{equation} 
of the depolarized scattered field, where the bracket indicates 
for an ergodic system
likewise a time or equilibrium ensemble average.

On assuming that the orientation of a particle is decoupled from the 
particles translation, $|G_{VH}(q,t)|$ can be factorized as 
\cite{Degiorgio:94,Degiorgio:95}
\begin{equation} \label{gvh}
   |G_{VH}(q,t)|=Nf^2(q)\frac{\beta^2}{15}F_s(q,t)F_r(t),
\end{equation}
where
\begin{equation}
   F_r(t)=4\pi\langle 
                 Y_{2,1}^*(\mbox{\boldmath$u$}_1(0)) 
                 Y_{2,1}(\mbox{\boldmath$u$}_1(t))
              \rangle
\end{equation}
is the rotational self--correlation function and
\begin{equation}
   F_s(q,t)=\langle 
               e^{i\mbox{\boldmath\scriptsize$q$}\cdot
                  [\mbox{\boldmath$\scriptstyle R$}_1(0)-
                   \mbox{\boldmath$\scriptstyle R$}_1(t)]}
            \rangle
\end{equation}
denotes the translational self--correlation function, also 
known in the literature
as the self--intermediate scattering function. The assumption 
of translational and rotational
decoupling which leads to Eq. (\ref{gvh}) can be shown in case of 
spherically symmetric interacting
particles to be rigorously true at short times, i.e. to first order 
in $t$ in a short--time
expansion based on the generalized Smoluchowski equation 
\cite{Jones:Pusey:91,Degiorgio:95}.
At longer times, the assumed decoupling is not strictly valid, 
but for hard sphere suspensions
there is at least experimental evidence that deviations are 
small for all times
\cite{Jones:Pusey:91,Degiorgio:95}.

If HI are neglected, $F_r(t)$ becomes exponential
\begin{equation} \label{freerotation}
   F_r(t)=e^{-6D^r_0t},
\end{equation}
where $D^r_0=k_BT/(8\pi\eta a^3)$ is the Stokesian rotational 
diffusion coefficient of a 
single particle of radius $a$, and $\eta$ is the shear 
viscosity of the suspending fluid.
Eq. (\ref{freerotation}) arises from the fact that the 
orientational diffusion of particles
with radial symmetric pair interactions are independent from 
each other if the HI
are not considered. In the case of non--interacting spherical particles,
$F_s(q,t)$ also reduces to an exponential function , i.e. \cite{Berne:Pecora}
\begin{equation}
   F_s(q,t)=e^{-q^2D^t_0t},
\end{equation}
where $D^t_0=k_BT/(6\pi\eta a)$ is the translational diffusion coefficient at
infinite dilution.

We focus now on the short--time behaviour of interacting colloidal particles. 
The
short--time rotational diffusion coefficient $D^r_s$ is 
defined by the short--time
behaviour of $F_r(t)$ as
\begin{equation} \label{drs}
   D^r_s=-\frac{1}{6}\lim_{t\to 0}\frac{\partial\ln F_r(t)}{\partial t}.
\end{equation}
This measurable quantity contains the configuration--averaged effects of the
HI on the short--time rotational diffusion of a spherical particle. 
As we will show in the
next section, $D^r_s$
crucially depends an system parameters like the volume fraction $\phi$, the
amount of added electrolyte and the effective particle charge.
One finds $D^r_s\rightarrow D^r_0$ in the limit $\phi\rightarrow 0$.

Notice that the short--time limit $t\rightarrow 0$ in Eq. (\ref{drs}) 
actually means that
$\tau^r_B\ll t\ll 1/D^r_0$, with $\tau^r_B=\Theta/(8\pi\eta a^3)$ 
being the relaxation time 
of the angular momentum of a colloidal sphere with moment of inertia 
$\Theta$. For
uniform spherical particles is $\tau^r_B\approx\tau^t_B$, where 
$\tau^t_B=m/(6\pi\eta a)$ is the momentum relaxation time of a 
particle with mass $m$ \cite{Jones:Pusey:91}.
It is $\tau^r_B\approx 10^{-8}s$ for typical suspensions. Most dynamic
light scattering experiments are restricted to correlation times
$t>10^{-6}s\gg\tau^r_B$. As a consequence, inertial effects arising from 
the linear and 
angular momentum relaxation of the particles are not resolved, so that in 
DDLS only
the relaxation of the particle orientations and positions is probed.
This fact allows for a coarse--grained description of Brownian motion 
on the basis
of the generalized Smoluchowski equation, which describes essentialy this 
relaxation \cite{Jones:Pusey:91,Jones:1:88}.
On this level of description, HI can be considered to act instantaneously.

We further note that $1/D^r_0=\tau_I/3$, where $\tau_I=\sigma^2/D^t_0$ is the 
socalled structural relaxation time, i.e. the time roughly needed 
for a non--negligible
change of the direct interactions due to configurational relaxation.
Typically one finds $\tau_I\approx 10^{-3}s$ such that the short--time
regime $\tau^r_B\ll t\ll\tau_I\approx 1/D^r_0$ is well seperated from 
the long--time
regime $t\gg\tau_I$. For the latter regime, memory effects are 
important and lead to
deviations in $|G_{VH}(q,t)|$ from purely exponential decay.

\section{Short--time rotational diffusion}
\label{diffusionsection}

For calculating $D^r_s$, it is necessary at first to specify the 
time evolution of
the suspended particles. Since we restrict ourselves to correlation times 
$t\gg\tau^r_B$, the suspension dynamics is properly described by the 
generalized
Smoluchowski equation.
In Smoluchowski dynamics, $F_r(t)$ can be expressed as
\begin{equation} \label{fr}
    F_r(t)=4\pi\langle 
                 Y_{2,1}^*(\mbox{\boldmath$u$}_1(0))
                 e^{\hat{O}^\dagger t}Y_{2,1}(\mbox{\boldmath$u$}_1(0))
              \rangle,
\end{equation}
where
\begin{eqnarray}
   \nonumber
       \hat{O}^\dagger =\sum_{i,j=1}^N
       & &\bigg\{
          \left(
             \nabla_i-\frac{1}{k_BT}\nabla_i\Phi
          \right)\cdot
          \left(\mbox{\boldmath$D$}^{tt}_{ij}\cdot\nabla_j+
                \mbox{\boldmath$D$}^{tr}_{ij}\cdot\hat{\mbox{\boldmath$L$}}_j
          \right)+
   \\ \label{backwardoperator}
       & &~~~
          \hat{\mbox{\boldmath$L$}}_i\cdot
          \left(\mbox{\boldmath$D$}^{rt}_{ij}\cdot\nabla_j+
                \mbox{\boldmath$D$}^{rr}_{ij}\cdot\hat{\mbox{\boldmath$L$}}_j
          \right) 
       \bigg\},
\end{eqnarray}
is the adjoint (or backward) Smoluchowski operator 
\cite{Jones:Pusey:91,Degiorgio:95,Jones:1:88}.
Here 
$
   \hat{\mbox{\boldmath$L$}}_j=
   \mbox{\boldmath$u$}_j\times\frac{\partial}{\partial\mbox{\boldmath$u$}_j}
$
is the gradient operator in the compact space of orientations of the 
$j$th particle, with 
$\mbox{\boldmath$u$}_j$ being the unit vector pointing in the 
direction of its optical axis.
In writing Eq. (\ref{backwardoperator}) it has been assumed that 
the total energy
$\Phi=\Phi(\mbox{\boldmath$R$}^N)$ of the $N$ particles depends 
only on the center--of--mass
positions 
$\mbox{\boldmath$R$}^N=(\mbox{\boldmath$R$}_1,\ldots\mbox{\boldmath$R$}_N)$ 
and
not on the particle orientations. We further assume $\Phi$ to be 
pairwise additive, i.e.
\begin{equation}
    \Phi(\mbox{\boldmath$R$}^N)=
         \frac{1}{2}\sum_{k,l=1}^N\,\!\!\!'\:u(R_{kl}),
\end{equation}
where $u(r)$ is a spherically symmetric pair potential, and 
$R_{kl}=|\mbox{\boldmath$R$}_{kl}|$ with 
$\mbox{\boldmath$R$}_{kl}=\mbox{\boldmath$R$}_k-\mbox{\boldmath$R$}_l$.
The prime indicates that the term 
$k=l$ is excluded from the sum. The diffusivity tensors
$\mbox{\boldmath$D$}^{ab}_{ij}$, with $a,b\in\{r,t\}$, embody the 
HI between the particles
by coupling the forces 
$\mbox{\boldmath$F$}^N=(\mbox{\boldmath$F$}_1,\ldots\mbox{\boldmath$F$}_N)$
and torques
$\mbox{\boldmath$T$}^N=(\mbox{\boldmath$T$}_1,\ldots\mbox{\boldmath$T$}_N)$
acting on the particles to their translational velocities
$\mbox{\boldmath$V$}^N=(\mbox{\boldmath$V$}_1,\ldots\mbox{\boldmath$V$}_N)$
and angular velocities 
$\mbox{\boldmath$\omega$}^N=(\mbox{\boldmath$\omega$}_1,
                       \ldots\mbox{\boldmath$\omega$}_N)$:
\begin{equation}
          {\mbox{\boldmath$V$}^N \choose \mbox{\boldmath$\omega$}^N}=
               \frac{1}{k_BT}
               {\mbox{\boldmath$D$}^{tt}(\mbox{\boldmath$R$}^N)  
               \mbox{\boldmath$D$}^{tr}(\mbox{\boldmath$R$}^N) \choose 
               \mbox{\boldmath$D$}^{rt}(\mbox{\boldmath$R$}^N)  
               \mbox{\boldmath$D$}^{rr}(\mbox{\boldmath$R$}^N)}
               {\mbox{\boldmath$F$}^N \choose \mbox{\boldmath$T$}^N}.
\end{equation}
For spherical particles, the diffusivity tensors depend only 
on the position vectors.
The tensors 
$\mbox{\boldmath$\mu$}^{ab}_{ij}=\mbox{\boldmath$D$}^{ab}_{ij}/(k_BT)$
are called mobility tensors.

Using Eqs. (\ref{drs}) and (\ref{fr}), one can express $D^r_s$ into the form
\begin{equation}
   D^r_s=D^r_0H^r_s,
\end{equation}
and the dimensionless diffusion coefficient $H^r_s$ is given by
\begin{equation}
     H^r_s=\frac{1}{3D^r_0}
         \langle
         \mbox{Tr}\mbox{\boldmath$D$}^{rr}_{11}(\mbox{\boldmath$R$}^N)\rangle.
\end{equation}
Here $\mbox{Tr}\mbox{\boldmath$D$}^{rr}_{11}$ denotes the sum on the 
diagonal elements of the
tensor $\mbox{\boldmath$D$}^{rr}_{11}$.
Due to the many--body nature of the HI, it is not possible to 
perform an exact evaluation
of $H^r_s$, valid at all particle concentrations. For small volume fractions 
however, when the
mean interparticle distance gets sufficiently large, it becomes possible 
to obtain a good
approximation for $D^r_s$ by considering only two--body and, to leading 
order, three--body
contributions to the HI.

By using a rooted cluster expansion, Degiorgio et.al. have shown that
the normalized short--time rotational diffusion coefficient can be 
expressed as 
a series \cite{Degiorgio:95}
\begin{equation}
   \label{hydrovirial}
   H^r_s=1+H^r_{s1}\phi+H^r_{s2}\phi^2+H^r_{s3}\phi^3+\ldots\   . 
\end{equation}
The coefficient $H^r_{s1}$ of the linear term is expressable in 
term of integrals involving only
hydrodynamic two--body contributions and the radial 
distribution function $g(r)$.
Explicitly, $H^r_{s1}$ is given by \cite{Degiorgio:95,Cichocki:88}
\begin{equation}
   \label{hrs1}
   H^r_{s1}=\int_\sigma^\infty dr~ r^2 
            g(r)8\pi\eta\left[\alpha^{rr}_{11}(r)+2\beta^{rr}_{11}(r)\right],
\end{equation} 
and this expression involves two scalar mobility functions
$\alpha^{rr}_{11}(r)$ and $\beta^{rr}_{11}(r)$ depending on the 
interparticle distance $r$.
These functions are calculable by means of a series expansion in 
even powers of $(a/r)$
\cite{Jones:Schmitz:88}.
We only quote the leading terms
\begin{eqnarray}
   \label{alpha}
      8\pi\eta a^3\alpha^{rr}_{11}(r)&=&-3\left(\frac{a}{r}\right)^8+
      {\mathcal{O}}\left(\left(\frac{a}{r}\right)^{10}\right)
   \\ \label{beta}
      8\pi\eta a^3\beta^{rr}_{11}(r)&=&
      -\frac{15}{4}\left(\frac{a}{r}\right)^6-
      \frac{39}{4}\left(\frac{a}{r}\right)^8+
      {\mathcal{O}}\left(\left(\frac{a}{r}\right)^{10}\right).
\end{eqnarray}
Inserting Eqs. (\ref{alpha}--\ref{beta}) in Eq. (\ref{hrs1}) leads to
\begin{equation}
   H^r_{s1}=\int_2^\infty dt~ g(t)\left[ 
            -\frac{15}{2}t^{-4}-\frac{45}{2}t^{-6}+
            {\mathcal{O}}\left(t^{-8}\right)\right]
\end{equation}
with $t=r/a$.

The second coefficient $H^r_{s2}$ in the expansion of 
Eq. (\ref{hydrovirial}) is for more
difficult to evaluate since it involves hydrodynamic 
three--body contributions.
Using the method of reflections, Jones (in Ref. \cite{Degiorgio:95}) 
was able to calculate the 
leading term in the far--field expansion of the irreducible 
three--body mobility tensor.
Considering only this term, $H^r_{s2}$ is approximated by a 
three--fold integral
\begin{equation} \label{hrs2}
   H^r_{s2}=\frac{225}{64}\int^1_0 dt_{12} \int^1_0 dt_{13} \int^1_{-1} d\xi_1
            g^{(3)}(t_{12},t_{13},\xi_1) f(t_{12},t_{13},\xi_1),
\end{equation}
which involves the static three--particle distribution function 
$g^{(3)}$ in 
dependence of
$t_{12}=2a/R_{12}$, $t_{13}=2a/R_{13}$, and 
$\xi=\mbox{\boldmath$R$}_{12}\cdot\mbox{\boldmath$R$}_{13}/(R_{12}R_{13})$.
The somewhat lengthy expression for the function $f(t_{12},t_{13},\xi_1)$ 
is given in Eqs.
(36--37) of Ref. \cite{Degiorgio:95} and will not be repeated here.

For dilute to moderately concentrated suspensions, it is sufficient 
to consider only
the leading three terms in the expansion Eq. (\ref{hydrovirial}), 
with $H^r_{s1}$ and
$H^r_{s2}$ expressed through Eqs. (\ref{hrs1}) and (\ref{hrs2}).
It should be stressed that Eq. (\ref{hydrovirial}) is, 
up to this point, not just a virial expansion of 
$H^r_s$ in powers of $\phi$, since $g(r)$ and 
$g^{(3)}(\mbox{\boldmath$r$,$r$'})$ are also $\phi$--dependend.
To proceed further, one needs to specify the static distribution
functions, which themselves depend crucially on the
form of the pair potential $u(r)$. It is at this point where the
important differences in the short--time behaviour between
charge--stabilized suspensions and suspensions of hard spheres
come from.

For later comparison consider first a dilute suspension of colloidal
hard spheres (typically with $\phi<0.1$).
It is then allowed to use in Eqs. (\ref{hrs1}) and (\ref{hrs2}) 
for consistency the leading order virial expansion of the static
distribution functions $g_{HS}(r)$ and 
$g^{(3)}_{HS}(\mbox{\boldmath$r$,$r$'})$
to first and zeroth order, respectivly. If the exact
numerical input for the two--body functions 
$\alpha^{rr}_{11}(r)$ and $\beta^{rr}_{11}(r)$ is employed, the following
truncated virial expansion for $H^r_s$ is obtained 
\cite{Degiorgio:95}
\begin{equation}
   \label{hardsphereresult}
   H^r_s=1-0.630\phi-0.67\phi^2+\mathcal{O}(\phi^3).
\end{equation}
The radial distribution function $g_{HS}(r)$ has its maximum at contact,
i.e. $g_{HS}(r=\sigma^+)>1$, and the contact value is a monotoneously
increasing function of the volume fraction in the liquid regime
$\phi\le 0.49$. In the case of hard spheres, it is therefore necessary
to include many terms in the expansion of the mobility tensors in powers
of $a/r$. In fact, if the two--body scalar functions in Eq. (\ref{hrs1})
are approximated by series expansions up to $\mathcal{O}(r^{-20})$,
we would get a value of $-0.578$ for the first virial coefficient,
significantly different from the exact numerical result $-0.630$.

\section{Calculation of $D^r_s$ for charge--stabilized particles}
\label{modelsection}

In the present work, we study the short--time rotational diffusion of
suspensions of charge--stabilized particles. It is crucial to note in
contrast to hard sphere suspensions that charge--stabilized systems at 
sufficiently low ionic strength exhibit 
even at
small volume fractions spatial correlations 
with pronounced oscillations in $g(r)$ (cf. Fig. \ref{gvonr}).
For this reason, it is in general not possible for 
charge--stabilized suspensions to use a virial expansion for the
static distribution functions. Consider for example the zeroth
density limit $g_0(r)=\exp\left[-\beta u(r)\right]$ of the radial
distribution function. Due to the long--range nature of the
electrostatic pair forces, extremely small values of $\phi$ are needed
for $g(r)$ to approach this limit.
As a consequence, $g(r)$ has to be calculated from appropriate
integral equation schemes or from time--consuming computer simulations.
In other words, charge--stabilized suspensions at small to
moderately large $\phi$ can be regarded
as dilute with respect to the suspension hydrodynamics
(which allows to use a rooted cluster expansion truncated after the 
three--body contributions), but they are ``concentrated'' as far as
the microstructure is concerned.

In performing the numerical integration in Eq. (\ref{hrs1}), we
have decided to calculate $g(r)$ using the rescaled mean spherical
approximation (RMSA) \cite{Naegele:88}, mainly
for its numerical simplicity but also since it has been found to be an
efficient fitting device of experimentally determined structure
factors \cite{Naegele:Habil:published}.

Our RMSA results for $g(r)$ are based on the effective macroion fluid
model of charge--stabilized suspensions. In this model, the effective
pair potential $u(r)$ between two charged particles is described as a
sum of a hard--sphere potential with diameter $\sigma$ and a screened Coulomb 
potential 
\begin{equation}
   \label{dlvo}
   \beta u(r)=
            L_B\left(\frac{Ze^{\kappa\sigma /2}}{1+\kappa\sigma /2}\right)^2
            \frac{e^{-\kappa r}}{r}, \mbox{\ \ } r>\sigma.
\end{equation} 
Here $\beta=1/(k_BT)$, 
$L_B=\beta e^2/\epsilon$ is the Bjerrum length, $e$ the elementary charge,
$\epsilon$ the dielectric constant of the suspending fluid, 
and $Z$ the effective charge (in units of $e$) of a colloidal
particle. 
The equation
\begin{equation}
   \kappa^2=4\pi L_B\left[ n|Z|+2n_s\right]
\end{equation}
defines the Debye--H\"uckel screening parameter $\kappa$,
where $n$ and $n_s$ are the number densities
of colloidal particles and of an added 1--1 electrolyte, respectivly.
It is assumed that the counterions are monovalent.

The leading--term expression Eq. (\ref{hrs2}) for $H^r_{s2}$ has the
triplet distribution function 
$g^{(3)}(\mbox{\boldmath$r$,$r$'})$
as input. There are, to the best of our knowledge, so far no
manageable numerical schemes available which provide decent approximations
for $g^{(3)}(\mbox{\boldmath$r$,$r$'})$ in the effective macroion fluid
model. To calculate the three--body contribution $H^r_{s2}$ given 
by Eq. (\ref{hrs2}),
we therefore use for simplicity Kirkwood's
superposition approximation \cite{Hansen:McDonald}
\begin{equation} \label{kirkwood}
    g^{(3)}(\mbox{\boldmath$r$},\mbox{\boldmath$r$}')=
                 g(|\mbox{\boldmath$r$}-\mbox{\boldmath$r$}'|) 
                 g(r)
                 g(r'),
\end{equation}
with $g(r)$ calculated again with the RMSA scheme. The three--fold
integration is performed using a Monte Carlo method.

We have stressed that in the case of hard spheres many terms need to be
considered in the far--field expansions of the hydrodynamic
mobility tensors, since configurations of nearly
touching hard spheres are very likely. The situation is quite different
for charge--stabilized suspensions: the strong repulsion of the
electric double layers keeps the particles apart from each other, 
resulting in a very small
probability of two or more spheres getting close
to each other. Indeed, the $g(r)$
for charged particles remains essentially zero for particle seperations
comparable to the Debye--H\"uckel screening length $\kappa^{-1}$
(cf. Fig. \ref{gvonr}).

Therefore, in charge--stabilized suspensions at sufficiently 
low ionic strength, it is possible to account only for the first
few terms in the far--field expansions of the mobility tensors.
In our calculations of $H^r_{s1}$, we include terms up to 
$\mathcal{O}\left((a/r)^{20}\right)$ in the expressions of the 
two--body mobility functions \cite{Jones:Schmitz:88}.
Our results will show that only the first three terms up to 
$\mathcal{O}\left(r^{-10}\right)$ are significantly contributing to 
$D^r_s$.

\section{Results and discussion}
\label{resultssection}

In this section we present our numerical results for the normalized
short--time rotational diffusion coefficient $H^r_s$ of
charge--stabilized suspensions as a function of volume fraction
$\phi$, effective charge $Z$, and concentration of added 1--1 
electrolyte $n_s$. The system parameters used in our calculations are typical 
for partially crystalline spherical particles made of a fluorinated
polymer and dispersed in an index--matching solvent
mixture of water with 20\% urea: $\epsilon=87.0$, $T=294$K, and
particle diameter $\sigma=900$\AA\ \cite{Bitzer:private}.
The effective charge chosen in our RMSA calculations of $g(r)$ is
$Z=500$ for those results where $Z$ is kept constant.
If not stressed differently, we have included in our
calculations of $H^r_s$ two--body
contributions up to order $r^{-20}$ together with the leading
three--body contribution.

In the following, we will show that the volume fraction dependence of
$H^r_s$ for charged particles is qualitativly different from that
observed for hard spheres. The differences are most pronounced for deionized
charge--stabilized suspensions, where essentially all excess electrolyte
has been removed.
Thus, we first concentrate on suspensions with $n_s=0$.

\subsection{Deionized suspensions}

Fig. \ref{functionalbehaviour} displays our results for 
$H^r_s=D^r_s/D^r_0$ as a function of $\phi$ (crosses).
For comparison, the corresponding result, Eq. (\ref{hardsphereresult}),
for hard sphere suspensions is also shown in the figure.
Evidently, the influence of the HI on $H^r_s$ is less pronounced for charged
particles than for uncharged ones.
We will later show that salt--free (i.e. deionized) suspensions
are particularly weakly influenced by HI.

A best fit of our results for $H^r_s$ in deionized suspensions gives
the interesting result
\begin{equation}
   \label{quadraticphidependence}
   H^r_s=1-a_r\phi^2; \mbox{\ \ } a_r\simeq 1.15,
\end{equation}
i.e. a quadratic $\phi$--dependence! This should be contrasted with the
corresponding expression Eq. (\ref{hardsphereresult}) for hard spheres, where
the linear term in $\phi$ gives the dominant contribution to 
$H^r_s$ if $\phi<0.15$.
The quadratic $\phi$--dependence in Eq. (\ref{quadraticphidependence})
is valid up to surprisingly large volume fractions, typically up to 
$\phi\simeq 0.25$. Moreover, the coefficient $a_r$ is found to be 
nearly independent of the particle charge, for values of $Z$ large enough
so that the hard core of the particles remains completely masked
by their electrostatic repulsion (i.e. typically for $Z>200$).
This fact is illustrated in Fig. \ref{chargedependence}, which shows 
results for $H^r_s(\phi)$ for various values of $Z$. All graphs
in this figure can be fitted by the functional form   
Eq. (\ref{quadraticphidependence}), with numerical values of $a_r$ close to
$1.15$.

\begin{figure}[ph]
       \epsfxsize=10cm
       \epsfysize=6.5cm
       \epsfbox{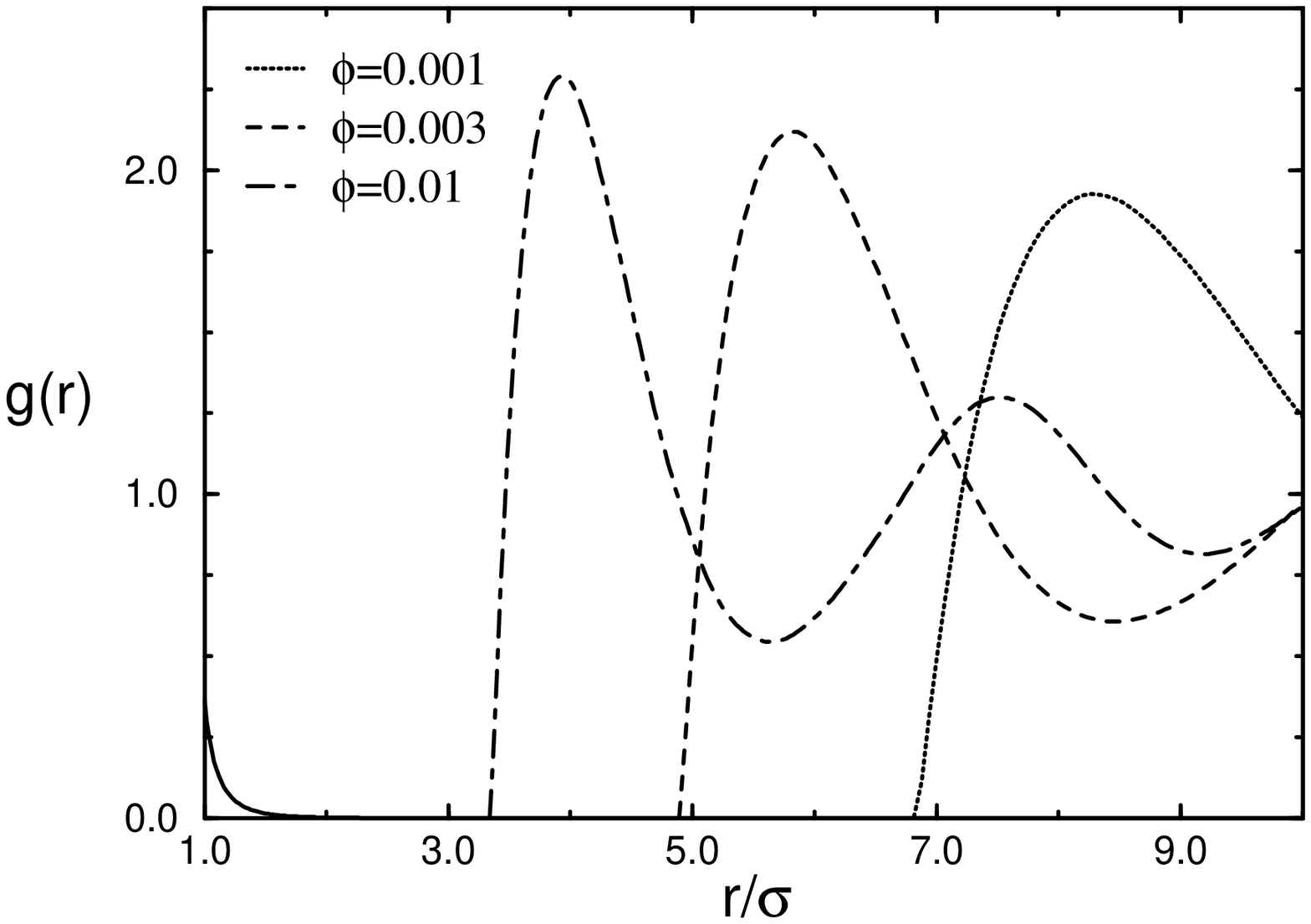}
       \caption{ \label{gvonr}
               RMSA results for the radial distribution function g(r), 
               modelling
               deionized suspensions at various volume fractions $\phi$, as
               indicated in the figure.
               System parameters are: $\sigma=900\mbox{\AA}$, $Z=500$,
               $\epsilon=87.0$, $T=294\mbox{K}$, and $n_s=0$.
               The full line is the graph of the two--body scalar mobility
               function
               $
                  -8\pi\eta a^3\left[
                                  \alpha^{rr}_{11}(r)+2\beta^{rr}_{11}(r)
                               \right]
               $, 
               including terms up to $\mathcal{O}(r^{-20})$
               (cf. Eq. (\ref{hrs1})).
               }
   \end{figure}

\begin{figure}[ph]
      \epsfxsize=10cm
      \epsfysize=6.5cm
      \epsfbox{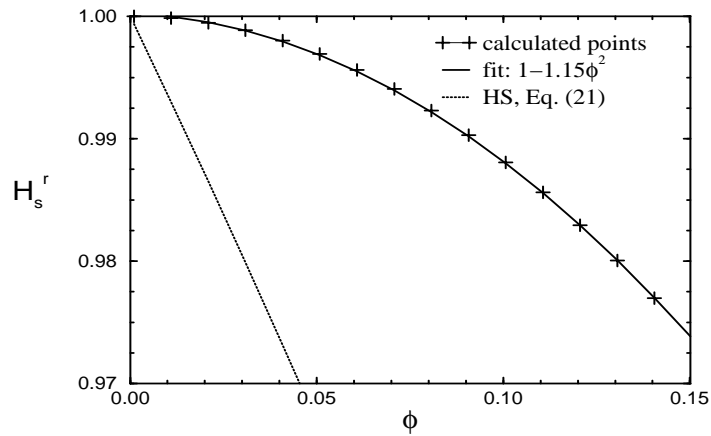}
      \caption{ \label{functionalbehaviour}
               Results for the normalized
               short--time rotational self--diffusion coefficient 
               $H^r_s=D^r_s/D^r_0$ for deionized systems, 
               in comparison with the
               corresponding hard--sphere result of Degiorgio et. al. [6].
               %
               Best fit has quadratic $\phi$--dependence:
               $H^r_s=1-1.15\phi^2$. System parameters as in Fig. \ref{gvonr}.
               }
   \end{figure}

\begin{figure}[ph]
      \epsfxsize=10cm
      \epsfysize=6.5cm
      \epsfbox{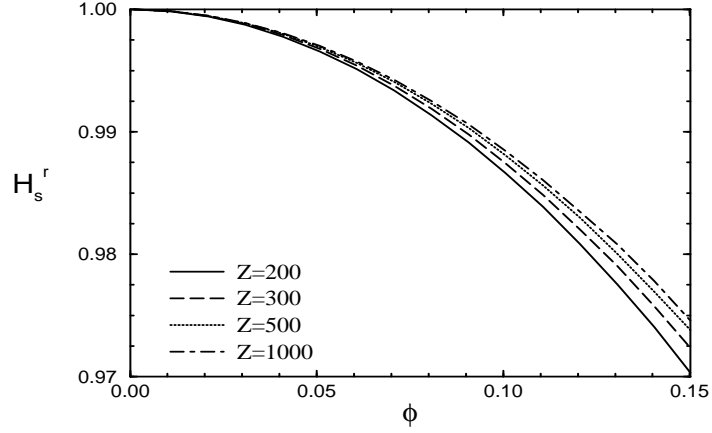}
      \caption{ \label{chargedependence}
               Normalized short--time rotational self--diffusion coefficient 
               $H^r_s$ for various 
               values of the effective charge  $Z$.
               All system parameters besides $Z$ as in Fig. \ref{gvonr}.
               }
   \end{figure}

A physical explanation for the, as compared to hard sphere suspensions,
weaker influence of the HI on $D^r_s$ in case of charged particles can be 
obtained from Fig. \ref{gvonr}. In this figure, RMSA results
for $g(r)$ are shown for different values of $\phi$. These graphs for 
$g(r)$ illustrate the pronounced interparticle corrrelations 
prevailing in deionized
suspensions down to very small volume fractions. 
Notice also from this figure that
the strong electrostatic repulsion gives rise to a 
``correlation hole'' centered around each particle, i.e. a spherical
region usually extending over several particle diameters with
(nearly) zero probability for finding another particle.
The size of this correlation hole increases with decreasing $\phi$.
As a result, there is at small $\phi$ only a weak hydrodynamic
coupling between the rotational motion of two or more spheres, and the
deviations of $D^r_s$ from its value $D^r_0$ at infinite dilution becomes
quite small. To quantify this point, we have plotted in Fig. \ref{gvonr}
the two--body scalar mobility function from Eq. (\ref{hrs1}), 
$
   -8\pi\eta a^3\left[
                   \alpha^{rr}_{11}(r)+2\beta^{rr}_{11}(r)
                \right],
$
versus the reduced interparticle distance.
According to Eqs. (\ref{alpha}--\ref{beta}) and Fig. \ref{gvonr}, 
this function 
is a rapidly decaying function of $r$. Obviously, the mobility 
function is very small at those
values of $r$ where $g(r)$ is different from zero, with the consequence 
that the value of the
integral in Eq. (\ref{hrs1}) is small indeed.

\begin{figure}[ph]
      \epsfxsize=10cm
      \epsfysize=6.5cm
      \epsfbox{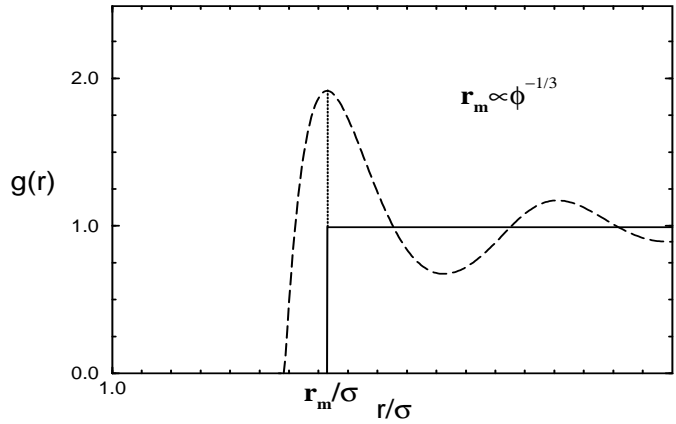}
      \caption{ \label{ehsdistribution}
               Typical RMSA--g(r) (dashed line), and corresponding
               radial distribution function $g_{EHS}(r)$ of the
               effective hard--sphere (EHS) model with $2a_{EHS}=r_m$.
               Note the scaling property 
               $r_m\propto\phi^{-1/3}$, valid for deionized suspensions.
               }
   \end{figure}

We present now an intuitive physical explanation for the quadratic 
$\phi$--de\-pen\-dence
of $D^r_s$, based on a characteristic scaling property of the principal peak
of $g(r)$, combined with a crude approximation for $g(r)$ which 
incorporates this
scaling property. In this approximation, the realistic $g(r)$
is replaced by a step function
\begin{equation}
   \label{gehs}
   g_{EHS}(r)=\Theta(r-2a_{EHS}), 
\end{equation}
where $a_{EHS}>a$ is an effective hard sphere radius, which accounts 
grosso modo
for the electrostatic repulsion between the particles. We refer to 
this simplified
description of the real pair structure as the effective hard sphere 
(EHS) model.
Using this model for $g(r)$ together with the far--field expansions 
Eqs. (\ref{alpha}--\ref{beta}) of the two--body mobility functions, it is
easy to calculate the leading terms in Eq. (\ref{hrs1}). The result is an
expansion
\begin{equation}
   \label{hrs1inehs}
   H^r_{s1}=-\frac{5}{16}x^{-3}-\frac{9}{64}x^{-5}-
            \frac{3}{64}x^{-7}+\mathcal{O}(x^{-9})
\end{equation}
in terms of the ratio of the effective radius to the actual particle 
radius, $x=a_{EHS}/a>1$.
At small $\phi$, one can neglect the three--body contribution
$H^r_{s2}\phi^2$ as compared to $H^r_{s1}\phi$, and $H^r_s$ 
is then approximated
in the EHS model as
\begin{equation}
   \label{hrsinehs}
   H^r_s=1-\left[\frac{5}{16}x^{-3}+\frac{9}{64}x^{-5}+\frac{3}{64}x^{-7}
            +\mathcal{O}(x^{-9})\right]\phi.
\end{equation} 
In the next step, we need to specify the effective radius. There are various
ways to estimate the magnitude of the effective radius in terms of 
the parameters of the
actual system \cite{Naegele:Habil:published}.
A reasonable choice adopted here and shown to be very useful in 
earlier applications of the
EHS model, is $2a_{EHS}=r_m$. Here $r_m$ is the position of the principal 
peak of the actual $g(r)$,
as calculated using the RMSA.
A typical RMSA--$g(r)$ and the corresponding EHS--$g(r)$ 
are depicted in Fig. \ref{ehsdistribution}.

For deionized suspensions with completely masked hard--core repulsion, 
it is well
known that $r_m$ and hence $x$ obeys the following scaling property 
in terms of $\phi$
\begin{equation}
   \label{xscaling}
    x\propto r_m\propto\phi^{-\frac{1}{3}}.
\end{equation}
This behaviour is due to the strong electrostatic repulsion, which leads to
$r_m\approx \bar{r}=a(4\pi/3\phi)^{-1/3}$, where $\bar{r}$ denotes the
geometrical average distance between neighboring particles.

In very diluted suspensions, $x$ becomes so large that it is sufficient to take
into account only the leading term in Eq. (\ref{hrsinehs}), 
which arises from the leading
term proportional to $r^{-6}$ in the expression Eq. (\ref{beta}) of
$\beta^{rr}_{11}(r)$.
The EHS model predicts for this case a quadratic $\phi$--dependence, since from
Eqs. (\ref{hrsinehs}) and (\ref{xscaling}) it follows
\begin{equation}
    H^r_s=1-\left[\frac{5}{16}x^{-3}+\mathcal{O}(x^{-5})\right]\phi
         =1-A\phi^2+\mathcal{O}(\phi^{8/3}),
\end{equation}
with $A$ determined to $A\simeq 0.60$ if $r_m$ is approximated by $\bar{r}$.
Hence we can conclude that the simple EHS model is 
for small $\phi$ in qualitative
agreement with our more refined numerical results for 
$D^r_s$ based on a more realistic $g(r)$.
The large deviation between $A$ and the coefficient $a_r$ in 
Eq. (\ref{quadraticphidependence})
arises from the "fine structure" in $g(r)$, not captured by the EHS model.

Since $r_m$ is determined by the volume fraction only, the 
EHS model further predicts 
$H^r_s$ to be independent of the particle charge, provided 
that $Z$ is large enough 
(i.e. $>200$) for $r_m\propto\phi^{-1/3}$ to be valid.
This prediction of the EHS model is again in good agreement with our numerical
finding for the $Z$--dependence of the coefficient $a_r$ in Eq.
(\ref{quadraticphidependence}).
That the two--body contribution to $H^r_s$, i.e. $H^r_{s1}$, 
is indeed not sensitive to 
changes in $Z$ follows from the observation that the principal peak of $g(r)$
becomes higher and narrower on increasing $Z$ at constant $\phi$, 
but its position
is nearly constant. The overlap region between $g(r)$ and 
$\left[\alpha^{rr}_{11}(r)+2\beta^{rr}_{11}(r)\right]$ which determines $a_r$ 
remains therefore also nearly constant.

\begin{figure}[ph]
      \epsfxsize=10cm
      \epsfysize=6.5cm
      \epsfbox{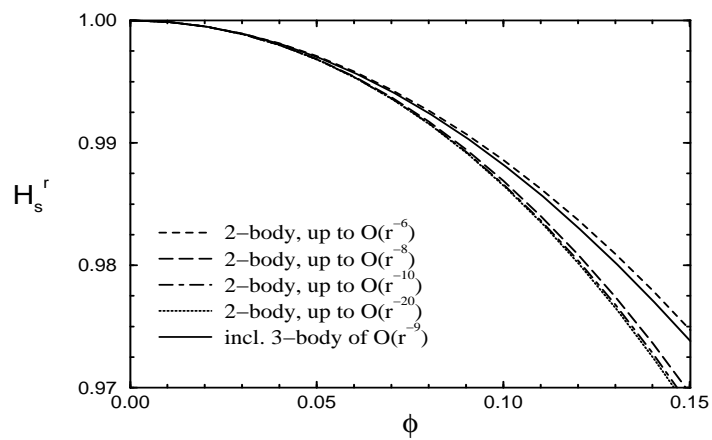}
      \caption{ \label{contributions}
               Dependence of $H^r_s$ on various two--body contributions, 
               and on the
               leading three--body contribution.  
               The various cases are indicated in the figure.
               Parameters as in Fig. \ref{gvonr}.
               }
   \end{figure}

The EHS model suggests that the quadratic $\phi$--dependence 
of $H^r_s$ in our numerical
results arises for small $\phi$ from the
leading term in the far--field expansion of the two--body mobility
function $\left[\alpha^{rr}_{11}(r)+2\beta^{rr}_{11}(r)\right]$.
For a numerical check of this assertion consider Fig. \ref{contributions},
which shows our result for $H^r_s$ if only the lowest order two--body
contribution of $\mathcal{O}(r^{-6})$ is considered (dashed line). 
In comparison we show the
result of the full calculation (full line), which was already displayed in Fig.
\ref{functionalbehaviour}, and which accounts for the leading
three--body term and all two--body contributions up to
$\mathcal{O}(r^{-20})$. We notice that both lines nearly
superimpose on each other even at larger volume fractions $\phi>0.1$, where the
three--body term and the higher order two--body terms are expected and 
found to give 
non--negligible contributions to $D^r_s$. However, 
at larger $\phi$ there is a fortuitous 
cancellation between the leading three--body contribution 
$H^r_{s2}\phi^2$ and the higher
order (i.e. $\mathcal{O}(r^{-8})$) two--body contributions to $H^r_{s1}\phi$,
which leaves the lowest order two--body contribution as the 
most significant term even
at larger $\phi$.

Fig. \ref{contributions} includes as the dotted line the result obtained for 
$H^r_s$ if the three--body contribution is neglected but all 
two--body contributions
up to $\mathcal{O}(r^{-20})$ are accounted for.
We obtain the same result if only two--body contributions up to 
$\mathcal{O}(r^{-10})$
are considered. Even the difference observed in the results for
$H^r_s$ including
two--body terms up to $\mathcal{O}(r^{-8})$ and $\mathcal{O}(r^{-10})$, 
respectively, 
is very small (cf. Fig. \ref{contributions}).  
As a conclusion, in charge--stabilized suspension it is 
justified to use a truncated
far--field expansion of the mobility functions even at larger $\phi$.

The fortuitous cancellation between the leading three--body term and the 
higher order two--body contributions observed in Fig. \ref{contributions}
could have been anticipated from the EHS model.
Using the first three terms in the expansion of Eq. (\ref{hrsinehs}), we
obtain the following expansion in $\phi$ by using Eq. (\ref{xscaling})
\begin{eqnarray}
   \nonumber
   H^r_{s1}\phi&=&\left[-\frac{5}{16}x^{-3}-\frac{9}{64}x^{-5}
                      -\frac{3}{64}x^{-7}
                      +\mathcal{O}(x^{-9})
                 \right]\phi
   \\
   \label{phiscalings}
               &=&-A\phi^2-B\phi^{\frac{8}{3}}-C\phi^{\frac{10}{3}}
                  +\mathcal{O}\left(\phi^{\frac{12}{3}}\right),
\end{eqnarray} 
with constants $A,B,C>0$. For an estimate of $B$ and $C$, let us
approximate $r_m$ again by $\bar{r}$. This leads to $B\simeq 0.41$ and
$C\simeq 0.21$. We do not need to consider higher
order terms in Eq. (\ref{phiscalings}), since these do not
contribute significantly to $H^r_{s1}$.

Eq. (\ref{hrs2}) for $H^r_{s2}$ yields together with Kirkwood's
superposition approximation, $g_{EHS}(r)$ 
(cf. Eq. (\ref{gehs})), and Eq. (\ref{xscaling}) the following 
result
\begin{equation}
   H^r_{s2}\phi^2=0.339~x^{-3}\phi^2=D\phi^3
\end{equation}
for the three--body contribution to $H^r_s$.
This result includes our finding 
\begin{equation}
   \left(H^r_{s2}\right)_{EHS}=x^{-3}\left(H^r_{s2}\right)_{HS},
\end{equation}
which states that the three--body contribution $\left(H^r_{s2}\right)_{EHS}$
is proportional to $\phi$.
The numerical value $\left(H^r_{s2}\right)_{HS}=0.339$ was first
obtained by Degiorgio et. al. \cite{Degiorgio:95} using
a Monto Carlo integration method.
Assuming $r_m=\bar{r}$, $D$ is determined as $D\simeq 0.65$.
It is now evident that there is a partial concellation between the
negative contribution
$\left[-B\phi^{8/3}-C\phi^{10/3}\right]$
and the positive contribution $D\phi^3$, leaving the term $-A\phi^2$
as the most significant contribution to $H^r_s$.

\subsection{Suspensions with added electrolyte}

\begin{figure}[ph]
      \epsfxsize=10cm
      \epsfysize=6.5cm
      \epsfbox{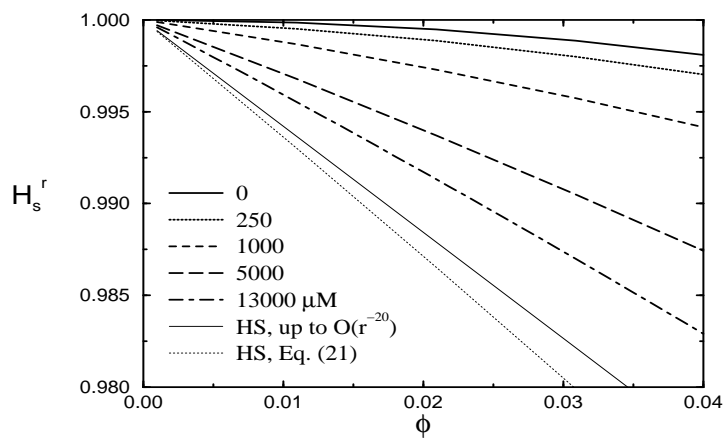}
      \caption{ \label{addedsalt}
               Volume fraction dependence of $H^r_s$ for various 
               amounts of added
               1--1 electrolyte, as indicated in the figure.
               System parameters as in Fig. \ref{gvonr}.
               }
   \end{figure}

We discuss now the dependence of $D^r_s/D^r_0$ on the amount od added
electrolyte.
In this context, it should be noted that most ``hard--sphere''
suspensions studied so far by DDLS are in fact suspensions of 
charged particles with a large amount of salt added to screen the
electrostatic repulsion (cf. e.g. \cite{Degiorgio:94,Degiorgio:95}).
Our results for $H^r_s(\phi)$ for various amounts of added 1--1 
electrolyte ranging from $n_s=0$ to $n_s=13mM$ are displayed in 
Fig. \ref{addedsalt}.
It is noted from this figure that addition of electrolyte leads
to a decrease of $H^r_s$, i.e. the rotational diffusion of the
particles becomes more affected by the HI with increasing
ionic strength. This is due to the fact that the system gradually
transforms with increasing $n_s$ into a hard--sphere--like dispersion. 
The behaviour of $H^r_s$ can be qualitatively explained on the basis
of the EHS model by noting for increasing $n_s$ that
the effective radius $a_{EHS}=r_m/2$ decreases and eventually
approaches the physical diameter $a$ for very large $n_s$.
It follows from Eq. (\ref{hrsinehs}) that $H^r_s$ becomes smaller
with decreasing $x\propto a_{EHS}$.

\begin{figure}[ph]
      \epsfxsize=10cm
      \epsfysize=6.5cm
      \epsfbox{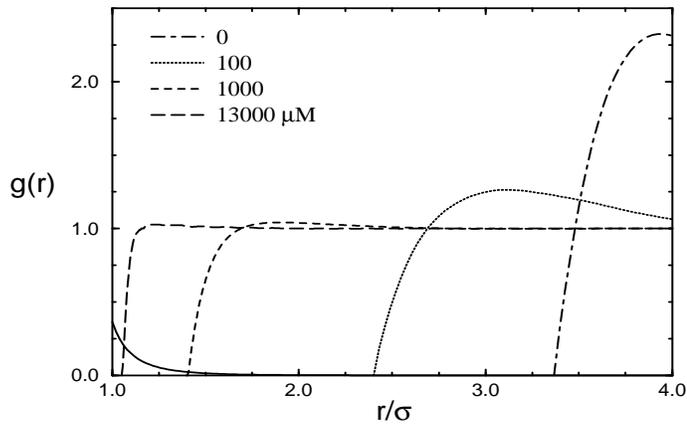}
      \caption{ \label{saltgvonr}
               RMSA results for g(r) for $\phi=0.01$, and for various
               amounts of added 1--1 electrolyte. Other parameters
               as in Fig. \ref{gvonr}.. 
               Full line: two--body mobility function
               $
                  -8\pi\eta a^3\left[
                                  \alpha^{rr}_{11}(r)+2\beta^{rr}_{11}(r)
                               \right]
               $ 
               as in Fig. \ref{gvonr}.            
               }
   \end{figure}

Going beyond the EHS model, we can learn more about the ionic strength
dependence of $H^r_s$ by considering Fig. \ref{saltgvonr}, which shows
RMSA results for $g(r)$ for a suspension of volume fraction 
$\phi=0.01$ and various amounts of added electrolyte.
The $g(r)$ of the deionized suspension has pronounced
oscillations, indicating strong particle correlations.
These oscillations become damped out as electrolyte is added, and the
peak position $r_m$ is shifted towards smaller values.
For the system under consideration, very large amounts of electrolyte
(i.e. $n_s\gg 13mM$) are needed to reach the hard--sphere limit,
where the RMSA--$g(r)$ becomes equal to the corresponding
Percus--Yevick solution \cite{Hansen:McDonald}.
Consider, e.g., the case $n_s=13mM$ in Fig. \ref{saltgvonr},
which corresponds to $\kappa\sigma\simeq 33$ and
$\beta u(r=\sigma^+)\simeq 6.0$.
The value of the pair potential at contact distance is still large
enough to create a small but non--negligible correlation hole
with zero probability of finding another particle.
In fact, it is well known 
\cite{Degiorgio:95,Jones:1:88,Cichocki:88}
that the value of $H^r_s$, and hence the
value of $D^r_s$, is extremely sensitive to the behaviour of the
radial distribution function near touching.
This is clearly seen from the fast decay of the two--body scalar
mobility function
$
   -8\pi\eta a^3\left[
                   \alpha^{rr}_{11}(r)+2\beta^{rr}_{11}(r)
                \right]
$
shown in Fig. \ref{saltgvonr}.

The scaling property $r_m\propto\phi^{-\frac{1}{3}}$
is invalid for larger amounts of electrolyte, when the particle diameter
becomes a second physically relevant length scale besides the
mean particle distance $\bar{r}$. The behaviour of $H^r_s(\phi)$
at small $\phi$ changes then gradually with increasing $n_s$ from 
a quadratic to a linear $\phi$--dependence.
At very large $n_s\gg 10mM$, $g(r)$ exhibits only tiny
changes in its form when $\phi$ is increased.
Hence $H^r_s$ gets more and more independent of $\phi$,
and $H^r_s(\phi)=1+H^r_{s1}(\phi)\phi+\mathcal{O}(\phi^2)$
becomes linear in $\phi$.
A deficiency of our calculations is the fact that we do not obtain the
exact hard--sphere result $\left(H^r_{s1}\right)_{HS}=-0.630$
in the limit $n_s\rightarrow\infty$. Instead the limiting
value $\left(H^r_{s1}\right)_{HS}=-0.578$ is reached, since we use
a series expansion up to terms of order $r^{-20}$ in calculating
$H^r_{s1}$.
However, this truncated expansion is sufficiently good for charge--stabilized
suspensions with moderate amounts of added electrolyte, where
$x>1.5$ holds (cf. \cite{Jones:1:88,Jones:Schmitz:88}).

\section{Conclusions}
\label{conclusionsection}

In the present work, we have 
investigated the combined effects of electrostatic interactions
and hydrodynamic interactions on the short--time rotational diffusion
coefficient $D^r_s$ in monodisperse suspensions of charge--stabilized
colloidal particles.

On the basis of the one--component macrofluid model, which neglects
electroviscous effects arising from the dynamic distortion of the
electric double layer around a particle, we have calculated 
$H^r_s=D^r_s/D^r_0$ for various systems as a function of volume
fraction, effective particle charge, and ionic strength.
For these calculations we have used a rooted cluster expansion derived
by Degiorgio et. al. \cite{Degiorgio:95}, combined with
far--field expansions of the two--body and three--body mobility
functions.
We have given evidence that this is a good and well founded approximation
in case of charge--stabilized suspensions, as long as the amount of added
electrolyte is not very large.

The short--time rotational diffusion of charged particles is less 
affected by the HI than the rotational diffusion of the uncharged ones.
For deionized suspensions of highly charged particles we have found a
quadratic volume fraction dependence of the form
$H^r_s=1-a_r\phi^2$, with a coefficient $a_r\simeq 1.15$, which is nearly
system independent.
The quadratic $\phi$--dependence extends up to rather large volume fractions
due to the fact that the leading three--body term is essentially
cancelled by the two--body contributions of $\mathcal{O}(r^{-8})$.
We have further shown that the qualitative behaviour of $H^r_s$ can
be understood in the framework of a simplified EHS model.

Incidently, dilute deionized suspensions are peculiar also with respect to
the $\phi$--dependence of the translational short--time self--diffusion
coefficient $D^t_s$. For this quantity, one obtains a non--analytic
concentration dependence of the form $D^t_s/D^t_0=1-a_t\phi^{4/3}$,
with $a_t\simeq 2.5$ \cite{Naegele:95:1}. The fractal
exponent $4/3$ follows also from the EHS model applied to translational
self--diffusion.

Upon addition of salt, the quadratic $\phi$--dependence of 
$D^r_s$ changes gradually to a $\phi$--dependence typical for hard spheres,
where the term linear in $\phi$ dominates for $\phi<0.1$.
Very large amounts of salt are required for the systems
investigated in this work to reach the hard--sphere limit. This
finding is due to the fast decay of the two--body mobility
functions with increasing $r$, which renders the two--body term
$H^r_{s1}$ extremly sensitive to distances close to the contact
distance of two particles.

Degiorgio et. al. \cite{Degiorgio:95} report a small
increase of the measured $D^r_s$ of hard spheres at the freezing
transition where $\phi_f=0.49$. The theoretical
argument to explain this enhancement is the slightly
increased free volume per particle in the crystal phase.
The resulting slightly increased interparticle distance turns the HI somewhat
less important. We expect that a similar effect should occur
for the freezing transition in charge--stabilized suspensions,
which occurs at much smaller $\phi$.
In our calculations we have used pair distribution functions $g(r)$
calculated with a method (RMSA) designed only for the fluid phase.
Therefore, we can not account for the effect of
the freezing transition on the short--time rotational diffusion.
Nevertheless, we anticipate that our calculations give reasonable
results even for the colloidal crystal phase, since our arguments
concerning the scaling property of the peak position $r_m$ of 
$g(r)$ should also apply to the crystal phase. Beside that we have shown that
$H^r_s$ is not very sensitive on the details of the shape of $g(r)$
near $r_m$. For these reasons, we expect only a small 
decrease of the coefficient $a_r$ at the freezing transition.
Therefore we have shown in this paper calculations of $H^r_s$
up to moderatly large $\phi$, although we know from the
Hansen--Verlet criterion that the systems studied here should be crystalline
already at small $\phi$.

To our knowledge no experimental or computer simulation data of
$D^r_s$ for deionized 
charge--stabilized suspensions have been published so far.
Because of the interesting differences in the rotational diffusion
between suspensions of charged and uncharged particles, it would
be worth to check the general predictions of our calculations against
experiments and computer simulations.
We finally mention that recent DDLS measurements of $D^r_s$ in
deionized suspensions of charged fluorinated polymer particles,
initiated by our work, compare
favorable with our calculations \cite{Bitzer:private}.

\vfill
\setcounter{totalnumber}{20}
\renewcommand{\textfraction}{0.0}
\renewcommand{\floatpagefraction}{1.0}
\renewcommand{\floatsep}{0mm}
\pagebreak
\section*{Figures}
\end{document}